# Evidences for higher nocturnal seismic activity at the Mt. Vesuvius

Adriano Mazzarella[1] and Nicola Scafetta[1]

[1]Meteorological Observatory, Department of Earth Sciences, Environment and Georesources
University of Naples Federico II, Largo S. Marcellino 10, 80138 Naples, Italy.

*Corresponding author: Adriano Mazzarella
Email: adriano.mazzarella@unina.it

**Abstract:**
We analyze hourly seismic data measured at the Osservatorio Vesuviano Ovest (OVO, 1972-2014) and at the Bunker Est (BKE, 1999-2014) stations on the Mt. Vesuvius. The OVO record is complete for seismic events with magnitude $M \geq 1.9$. We demonstrate that before 1996 this record presents a daily oscillation that nearly vanishes afterwards. To determine whether a daily oscillation exists in the seismic activity of the Mt. Vesuvius, we use the higher quality BKE record that is complete for seismic events with magnitude $M \geq 0.2$. We demonstrate that BKE confirms that the seismic activity at the Mt. Vesuvius is higher during nighttime than during day-time. The amplitude of the daily oscillation is enhanced during summer and damped during winter. We speculate possible links with the cooling/warming diurnal cycle of the volcanic edifice, with external geomagnetic field and with magnetostriction that should also stress the rocks. We find that the amplitude of the seismic daily cycle changes in time and has been increasing since 2008. Finally, we propose a seismic activity index to monitor the 24-hour oscillation that could be used to complement other methodologies currently adopted to determine the seismic status of the volcano and to prevent the relative hazard.

# 1. Introduction

The Mt. Vesuvius is one of the best studied volcanoes in the world. The 79 A.D. eruption, which buried the Roman city of Pompei and Ercolano, is worldwide famous. Since the 1631 subplinian eruption, the volcano has exhibited Strombolian activity and effusive-explosive eruptions alternating with periods of quiescence. The 1944 eruption ended the most recent volcanic cycle and since then, the volcano has been in a quiescent stage with low–energy seismic activity (cf. Scandone and Giacomelli, 2014; Gregori, 1996).

The Vesuvian area is interesting because it is still exposed to high volcanic and seismic hazard. This region is also important because of its highly dense urban settlements surrounding the Neapolitan area, which count more than 2 million inhabitants (Barnes, 2011). In fact, the Mt. Vesuvius is in a self-organized critical state in which a small perturbation can start a chain reaction that could lead to a catastrophe (c.f.: Luongo et al., 1996, 2000; Luongo and Mazzarella, 2001, 2003; De Falco et al., 2002).

Luongo et al. (1996) and De Falco et al. (2002) forecast that a violent type of eruption may occur during the 21st century. Scafetta and Mazzarella (2015) have analyzed the world wide historical earthquake database and found that the earthquake activity follows quasi 20- and 60-year cyclical patterns that are found also in numerous climatic data and showed that the worldwide seismic activity could increase during the next decade. Bragato (2015) found a quasi 60-year oscillation also in the Vesuvian eruptions using a 1631-1944 catalog. These results suggest the existence of predictable patterns of seismic activity that could be linked to volcano eruptions. If these patterns are better identified, their dynamical evolution could be used to evaluate the status of the volcano and to develop policies to alert the population and to mitigate the relative hazard. Thus, it is essential to understand the seismic-volcanic activity of the Mt. Vesuvius and its generating mechanisms during quiescent stages through a careful chemical, physical and statistical monitoring.

There have been several studies suggesting that seismic and volcano activity could be partially regulated by lunar and solar tidal forcing, which present diurnal and semidiurnal oscillations (e.g.: Kilston and Knopoff, 1983; Riguzzi et al., 2010; Gregori, 2015; etc.). However, Neuberg (2000) noted that the spectral peaks of seismic tremor correspond to exactly 24 and 12 hour periods that do not match the theoretical tidal spectrum that, on the contrary, shows several spectral peaks in the diurnal (waves: $K_1$, $O_1$, etc.) and the semidiurnal (waves: $M_2$, $S_2$, $N_2$, etc.) bands: a list of the main tidal constituents can be found in Palumbo and Mazzarella (1982). Thus, Neuberg (2000) proposed that meteorological parameters such as temperature, barometric pressure, wind and rain, which show the same diurnal and semidiurnal modulation of seismicity, should be considered possible candidates for a possible external modulation of earthquake and volcanic activity.

The 24-hour periodic behavior of seismic activity has been controversial. Shimshony (1970) analyzed only a 3-year worldwide NOAA catalog from 1968 to 1970 and claimed that it was characterized by a diurnal cycle with maxima during the nighttime. However, Knopoff and Gardner (1972) and Davies (1973) harshly critiqued Shimshony claiming that his result was an artifact due to the too low quality and shortness of the analyzed record. In addition, they claimed that cultural noise could bias the catalog in favor of reporting more events during local nights because during day-times the anthropogenic noise could mask natural shocks.

Duma and Vilardo (1998) analyzed the seismic activity recorded at the Osservatorio Vesuviano Ovest (OVO) station from 1972 to 1996 and showed by visual inspection alone that this activity could be characterized by a diurnal oscillation with maxima during nighttime. The OVO record, available from 1972 to 2014, is complete for events with magnitude $M \geq 1.9$. Based on their results, Duma and Vilardo speculated a possible relation linking the variations of solar activity and of the Earth's magnetic field to the Mt. Vesuvius seismic activity. A similar diurnal cycle was also claimed to exist in E-China, California and Austria (Duma and Ruzhin, 2003).

Herein we will show that the conclusion of Duma and Vilardo (1998) should be questioned because since 1997 the OVO record has not showed a significant daily oscillation any more. Thus, the finding of a daily oscillation in the Mt. Vesuvius seismic activity could have been accidental and/or restricted only to the period from 1972 to 1996. Consequently, the issue needs to be investigated again to solve the ambiguity.

We demonstrate that the OVO record misses the daily oscillation since 1997 because of its too low quality. We reached this conclusion by analyzing the alternative and higher quality seismic record that has been collected at the Bunker Est (BKE) observatory since 1999 (http://www.ov.ingv.it). The BKE record is considered to be of higher quality than the OVO record because it is complete for magnitude $M \geq 0.2$.

Our study of the BKE record confirms the existence of a persistent daily oscillation in the seismic activity of Mt. Vesuvius. We discuss possible links with the daily meteorological cycle, which also induces a correspondent thermal oscillation in rocks, with external geomagnetic field and with electro-magnetostriction processes.

Finally, because the significance of the observed pattern varies in time, we propose a simple methodology to monitor the evolution of the daily oscillation to determine the present status of the volcano and to prevent a possible hazard.

To characterize the diurnal cycle of the Vesuvius seismic activity, we have considered two different indexes: the count of seismic events with magnitude above the completeness threshold and the seismic total energy release, which is less sensitive to completeness problems (Kanamori and Anderson, 1975; Luongo et al., 1996). Our preliminary tests confirmed that the count and energy indexes provide similar qualitative results about the existence of a daily oscillation due to the well-known Gutenberg–Richter law expressing the relationship between the magnitude and total number of earthquakes in any given region (Gutenberg, Richter, 1954). However, the emergence of a daily oscillation is clearer when the seismic count event index is used (cf.: Duma and Vilardo, 1998). This indicates that the daily pattern is associated mostly to minor tremors, which are more numerous: in fact, minor tremors can be more easily and immediately induced by a daily oscillating forcing. On the contrary, the energy indexes stress the time occurrence of major events that are more likely the result of an accumulated stress. An accumulated stress can cause a major rupture at a more random time and, therefore, major seismic events are less likely to follow the timing of an external forcing on a hourly/daily scale. Herein we will study the time evolution of the count index. The monitoring of the frequency of minor tremors is particularly important because such events could be foreshocks of larger events. Further research will address the relation between the count and energy-based seismic indexes.

## 2. Data and analysis techniques

The OVO hourly seismic catalog (see link: http://www.ov.ingv.it/ov/doc/ctmag72_14gmt.txt) collects 11804 seismic events from February 1972 to September 2014 with magnitude between -0.4 and 3.6: see Figure 1A. The catalog is complete for magnitude $M \geq 1.9$ (Luongo et al., 1996; Duma and Vilardo, 1998; Duma and Ruzhin, 2003) for which there are only 1490 events. From 1972 to 1996, that is the period analyzed by Duma and Vilardo (1998), there are 1202 events with magnitude $M \geq 1.9$. However, from 1997 to 2014 there are only 287 events with magnitude $M \geq 1.9$.

The BKE hourly seismic catalog (see link: http://www.ov.ingv.it/ov/doc/ct_BKE_14.txt) collects 11042 seismic events from January 1999 to September 2014 with magnitude between -2.5 and 3.6: see Figure 1B. The catalog is complete for magnitude $M \geq 0.2$ (Orazi et al., 2013) for which there are 5035 events. Thus, since 1999 the BKE record should permit a more robust statistical analysis respect to the OVO record. The BKE record is considered the most reliable seismic record on the Vesuvius.

To reject the randomness of the seismic frequency, we adopted the two methodologies used by Knopoff and Gardner (1972) and Davies (1973). We observe that Duma and Vilardo (1998) did not perform a statistical analysis of their result but limited it to a visual inspection of hourly histograms.

The first method computes the hourly frequency distribution of seismic events during the 24-hour daily cycle. The result is tested against the null hypothesis that seismic events occur uniformly at any time of the day using a $\chi^2$ test with 23 degrees of freedom (df). The hypothesis is then rejected at the 95% (99%) level of confidence if the observed $\chi^2$ is found to be greater than 35.2 (41.6). The second method compares the day-time (from 6:00 am to 6:00 pm) and the nighttime (from 6:00 pm to 6:00 am) seismic activity. The result is then tested against the null hypothesis that the seismic events are equally distributed during the daytime and nighttime using a $\chi^2$ test with 1 df. The hypothesis is then rejected at the 95% (99%) level of confidence if the observed $\chi^2$ is greater than 3.84 (6.63).

The $\chi^2$ test is based on the evaluation of the expression:

$$\chi^2 = \sum_{i=1}^{df+1} \frac{(observed_i - expected)^2}{expected}, \qquad (1)$$

where df = 23 if the hourly frequency is used and df = 1 if the day-night test is used. The "expected" values are given by the observed average values in the two cases.

Moreover, to test whether the record presents a 24-hour oscillation, we also use the Chapman-Miller (CM) method that is commonly used in geophysics to identify solar, lunar diurnal and semidiurnal harmonics (Chapman and Miller, 1940; Malin and Chapman, 1970; Cecere et al., 1981). In the CM method, the seismic events are arranged in 12 groups according to lunar age. The group-sum sequences are harmonically analyzed and amplitudes and phase angles of investigated harmonics are obtained. The reliability of each harmonic is computed according to the vector probable error (vpe) equal to the root-mean-square radius of the point cloud distributed on the relative harmonic dial. The confidence level of the harmonic is computed according to the equation: $1 - \exp[-(0.833\ A/vpe)^2]$ and each harmonic is found to be confident at 95% (99%) level when the amplitude A $\geq$ 2.08 vpe (A$\geq$3.00 vpe). Here, the CM method is used only to investigate the 24-hour harmonic whose diurnal amplitude is expressed as percentage of the number of shocks with respect to the observed mean hourly value.

To test the temporal persistency of the investigated daily pattern in the Mt. Vesuvius seismic activity, we study: 1) the entire BKE record from 1999 to 2014; 2) the entire record for M ≥ 0.2 for which the catalog is complete to reduce a possible influence of cultural noise;  3) three contiguous sub-catalogs, each containing one third of the entire record. We also investigate also the seasonal modulation of the diurnal cycle according to Lloyds season subdivision (D months: November to February. E months: March, April, September, October. J months: May to August).

Finally, we propose a continuous index that measures the average daily oscillation of seismic activity based on a 12-month and 4-month moving intervals, respectively, that can be used to monitor the volcano activity.

## 3. The OVO seismic record

Figure 2A depicts the hourly frequency of the seismic events referring to the OVO catalog for M ≥ 1.9 from 1972 to 1996, which is the period analyzed by Duma and Vilardo (1998). Here there are 1202 events.  The original OVO record is given in Greenwich Mean Time (GMT), but here we transformed GMT into local time (LT) because we are interested in investigating the effective diurnal oscillation.  Figure 2B depicts the hourly frequency of the seismic events referring to the OVO catalog for M ≥ 1.9 from 1997 to 2014, which is the period that could not be analyzed by Duma and Vilardo (1998). Here there are only 287 events.  The red curve in the two figures is a data fit with a 24-hour sinusoidal function. The figures report the amplitude A, the hour of maximum occurrence $\phi$, and the average value $\mu$. The gray shadow refers to the nighttime from 6:00 pm to 6:00 am LT.

Figure 2 shows that from 1972 to 1996 the OVO record is characterized by an evident diurnal oscillation with maxima during the nighttimes, as Duma and Vilardo (1998) found. However, since 1997 the diurnal oscillation is very weakened.  From 1972 to 1996 we found $\chi^2$ = 243 with 23 degrees of freedom and the null hypothesis is then rejected at the 99% level of confidence because the observed $\chi^2$ is found to be greater than 41.6.  From 1997 to 2014 we found $\chi^2$ = 40.6 with 23 degrees of freedom. In this case the null hypothesis is accepted at the 99% level but rejected at 95% level.

To investigate better the issue, Figure 3 depicts the hourly frequency of the seismic events referring to the OVO catalog for M ≥ 1.9 from 1972 to 2014 for each year.

The figure shows 43 contiguous histograms (the gray shadow refers to the nighttime from 6:00 pm to 6:00 am LT).

Figure 3 shows that in the most of the years from 1972 to 1996 OVO recorded more seismic activity during nighttime. However, since 1997 only in 1999, 2000 and 2001, the OVO record registered more seismic activity during nighttime. Many other years are characterized by a too low number of events to detect a significant daily oscillation.

## 4. The BKE seismic record

Figure 4A depicts the hourly frequency of the seismic events referring to the entire BKE catalog containing 11804 events from 1/1/1999 to 7/9/2014. Figure 4B depicts the hourly frequency of the seismic events referring to the BKE catalog considered to be complete for M $\geq 0.2$ and containing 5035 events from 1999 to 2014. The original BKE time record is in GMT, but here we have transformed this variable into LT, as done above.

Figure 5A shows three panels depicting the hourly daily frequency of the entire records divided into three time-contiguous subsets containing 3935 (1/1/1999 - 14/8/2001), 3935 (14/8/2001 - 7/11/2008) and 3934 (8/11/2008 - 7/9/2014) events respectively. Figure 5B shows three panels depicting hourly daily frequency of the same seismic record analyzed as above but with M $\geq 0.2$.

Figure 6A shows three panels depicting the hourly daily frequency of the entire BKE catalog subdivided into the three Lloyds seasons D, E and J. Figure 6B shows three panels depicting hourly daily frequency of the BKE catalog for M $\geq 0.2$ and subdivided into the three Lloyds seasons D, E and J. Figures 7A and 7B are like Figure 6A and 6B, but they refer only to the 8/11/2008 - 7/9/2014 interval containing 3934 raw events and 1655 events with M $\geq 0.2$, respectively.

The red curve in the figures is a histogram fit with a 24-hour sinusoidal function. The figures report the amplitude A, the hour of maximum occurrence $\phi$, and the average value $\mu$. A visual inspection of Figures 4-7 clearly suggests the existence of a diurnal oscillation in the BKE record in all analyzed situations. The strongest amplitude of the cyclical pattern is shown in Figure 7 panel J, which refers to the summer Lloyds months from 8/11/2008 to 7/9/2014. The maximum of the events occurs always around midnight and the minimum around noon. The last 3934 earthquake sequence and the last 1655 earthquake sequence with M $\geq 0.2$ depicted in Figure 7 show the highest amplitude in respect to the obtained sub-catalog.

The $\chi^2$ results, reported in table 1, show that the null hypothesis is rejected at 99% both for the hourly frequency and for the day-night method, respectively. In BKE the nighttime seismic activity is statistically higher than the daytime one. The only ambiguous case refers to the "first third, M ≥ 0.2" record depicted in Figure 2B-top.

Figures 8A and 8B show the sequences of 24-hour histograms of the seismic data measured at the BKE Vesuvius station for each of the sixteen years from 1999 to 2014. The histograms are plotted contiguously for visual inspection and the gray areas represent the nighttime from 6:00 pm to 6:00 am LT. The total record and the record for M ≥ 0.2 are shown, respectively. This graphical representation suggests that every year the records do present a daily oscillation with maxima during the nighttime. Because the daily oscillation repeats every year, the hypothesis of randomness is further excluded.

Figures 9A, 9B and 9C are like Figure 8 related to the three Lloyds seasons: D, E and J. The daily cycle is confirmed, although it appears more evident since 2008 and in the E and J months.

Figures 8 and 9 show that from 1999 to 2001 the 24-hour pattern is more confused. We notice that during such a period numerous earthquakes with M ≥ 3.0 occurred whose aftershocks might have disrupted the daily cyclical pattern.

In order to statistically identify the diurnal oscillation in the occurrence of seismic events on Vesuvius, we also applied the Chapman-Miller method and the results are reported in Tables 2-7, which refer to the cases depicted in the figures. The tables report: 1) the amplitude (A) expressed in percentage relative to the mean value; 2) the vector probable error (vpe); 3) the phase in degree (°); 4) maximum of occurrence in hour (h); 5) the hourly mean; 6) the ratio A/vpe with the confidence level of the investigated daily harmonic. The results show that a daily oscillation is confirmed at a confidence level greater than 99% in all cases. Again, the only exception refers to the data shown in Figure 7 panel J where the confidence level is scarcely at 95%.

The above results suggest the adoption of a comprehensive index to measure the 24-hour cyclic seismic activity of the Mt. Vesuvius, which should be based on a record complete for events with magnitude M ≥ 0.2. This index is proposed in Figure 10 in two versions based on 12-month and 4-month moving intervals.

Figure 10A compares the number of seismic events for M ≥ 0.2 occurred during nighttime (6:00 pm - 6:00 am LT) and during daytime (6:00 pm - 6:00 am LT) for 12-month intervals. Figure 10B does the same for 4-month intervals. The intervals move by 1-month step. Figure 10C depicts the $\chi^2$ curves relative to each interval in the two cases.

Figure 10 fully summarizes the daily cyclical activity of the Mt. Vesuvius. On an annual scale the number of nighttime seismic events is always larger than during daytime since 1999. The same is typically observed also on a 4-month scale. In both cases, the difference between daytime and nighttime before 2002 is smaller than after 2007, as Figure 10C clearly shows, despite the number of events is actually larger before 2001. The $\chi^2$ curve referring to the 4-month moving intervals depicted in Figure 10 also highlights the existence of a strong annual oscillation with maxima during summer times and minima during winter times in particular after 2011 while this pattern is not equally evident from 1999 to 2011 but in 2002.

## 4. Discussion

Inspired by the work of Duma and Vilardo (1998), we have analyzed the daily oscillation in the seismic activity of the Mt. Vesuvius. As demonstrated above, we found that while the OVO record presents a daily oscillation from 1972 to 1996, as highlighted in Duma and Vilardo (1998), this oscillation nearly vanished afterward. This result yields to two main conclusions: (1) the finding of Duma and Vilardo (1998) about the existence of this daily oscillation could be questioned; (2) if a daily oscillation in seismic activity exists and it is important about the activity of the volcano and its forcing mechanisms, the OVO index would be inadequate for monitoring it. Thus, the issue needed to be reexamined using a more reliable seismic record such as BKE, which is available since 1999.

Our analysis of the BKE record has strongly highlighted the existence of a 24-hour oscillation. The seismic events are more frequent during the nighttime than during the day-time with a confidence level greater than 99%. The index shown in Figure 10 summarizes the dynamical behavior of the daily oscillation in the seismic activity and could be monitored to determine the status of the volcano and for hazard prevention.

Because the result is obtained also when only the $M \geq 0.2$ events are used, for which the catalog is complete, it is easier to reject the hypothesis that this daily cyclical pattern is simply an artifact due to cultural noise. Additional arguments are discussed below.

Paparo et al. (2003) measured the acoustic emission (AE) inside the Gran Sasso and in the Vesuvius areas and found a large diurnal oscillation with a maximum during nighttime. This result could be compatible with our result that during night the earthquake events are more frequent. Paparo et al. (2003) interpreted their result claiming that the daily warming-cooling cycle of rocks makes their external layer contract over a warmer and temporarily more expanded interior yielding to ruptures. This hypothesis would also explain why for numerous years the daily oscillation is less prominent during the winter time when the daily temperature excursion is the lowest. Paparo et al. (2004) also showed that the Mt. Vesuvius alternates periods of increasing pressure of its endogenous hot fluids and periods of decreasing pressure. This pattern could also have a daily recurrence related to the thermal cycle.

When some comparably stronger shock occurs, the crust experiences a resetting period and some past accumulation of elastic energy is being released. In this situation the diurnal cooling/heating cycle plays a necessarily secondary role. This is perfectly consistent with the findings of this paper where we noted that, from 1999 to 2001, the 24-hour cyclical pattern is nearly vanishes because during such a period numerous earthquakes with $M \geq 3.0$ occurred whose aftershocks might have disrupted it.

Recent observations have also linked the regular diurnal variations of the Earth's magnetic field, commonly known as "magnetic quiet-day solar daily variations" (Sq-variations) with the diurnal distribution of earthquakes. Duma and Ruzhin (2003) analyzed this geodynamic process of changing earthquake activity with the time of day, and found correlation in the daily distribution of earthquakes and Sq-variations in several region of the Earth. Duma's (2006) presentation deals with details of the electromagnetic model. His model builds on the mechanic Lorentz forces and relative torques which are generated by the regularly induced telluric currents in the Earth's lithosphere in presence of the Earth's main magnetic field.

A key effect ought to be related to planetary-scale stray electric currents (Foucault currents) induced by the solar wind (and channeled all over the globe) into the crust and lithosphere. These stray currents originate mechanical forces and torques (on the planetary scale) that generate conspicuous stress, which is more or less uniformly distributed all over the Earth. By this, they do contribute also to the sum of crustal stress that – eventually and in suitable environmental circumstances – can trigger an earthquake. Since the geomagnetic diurnal variation (including everyone of its components) is directly related to the electromagnetic induction effect by the solar wind, the geomagnetic variation should display a correlation with seismic activity (cf. Gregori and Lanzerotti, 1980).

To test such an electromagnetic hypothesis, we noted that Palumbo (1981) investigated the daily variations of the geomagnetic field at the available stations nearest stations to Vesuvius, namely at L'Aquila (200 km apart) and at Capri (40 km apart). Thus, we compare the 24-hour harmonics of geomagnetic field (Palumbo, 1981) with the 24-hour harmonics in the Vesuvius seismic activity to identify a possible association.

Palumbo's results, reported in Table 8, show a 24-hour coherence only with the 24-hour oscillation of the vertical (Z) component of the measured geomagnetic field averaged for all the available years and for each of the three Lloyds seasons (D, E and J). In addition to the well known daily oscillation, Table 8 highlights that the amplitude of the daily cycle in Z is lower in winter and its maximum occurs just before midnight.

In order to further confirm such a pattern, we also analyze the Z component of the geomagnetic field measured in several locations in Spain (Cueto et al., 2003) and in Turkey (Celik et al., 2012) that are approximately at the same latitude of the Mt. Vesuvius. The results are reported in Table 9.

The comparison with the seismic results shows a strong correlation between the seismic activity on the Mt. Vesuvius and the Z component of the geomagnetic field at the same latitude. In fact, in both cases we observe a maximum during the nighttime and an increase of the effect from the winter to the summer. Also the timing of the 24-hour cycle, with the geomagnetic one occurring just before the seismic one reveals a plausible casual relations.
Thus, we conclude that the seismic activity could be triggered also by the strength of the Z component of the geomagnetic field because such a component should be associated to a horizontal component of the Lorentz force that stresses the crust.

Electrostriction effects (a property of all electrical non-conductors, or dielectrics, that causes them to change their shape under the application of an electric field) could also favor seismic activity in the presence of a varying electromagnetic field. However, this phenomenon could provide only a secondary contribution because appears to be strictly localized. The Lorentz force during night should also be enhanced by the reduction of the electric field (cf.: Paparo et al., 2003) that could imply larger electric currents in the ground. Essentially, during daytime a charging process of the ground occurs, followed by a ground electric current discharge during the night.

Mazzarella and Palumbo (1988) analyzed the catalog of Italian earthquakes and found that the diurnal components of the geomagnetic field may induce mechanical stresses in the crust through a magnetostriction process that causes ferromagnetic materials (iron, nickel) to change their shape and dimensions when they are kept in a magnetic field. Such a process has a daily oscillation because triggered by the vertical component of the geomagnetic field which, as seen above, is maximum during nighttime. In addition, magnetostriction processes depends also by the temperature.

Another evidence for a possible link between seismic activity and the geomagnetic field can be observed by comparing Figure 3 and Figure 8. Here it is is evident that the number of seismic events increases every about 10 years. Periods of maximum activity occur in 1978-1982, 1989-1991, 1999-2001, 2009-2013. This quasi 11-year seismic oscillation appears coherent with the 11-year solar oscillation.

By taking into account the above discussion, the fact that the detected seismic daily oscillation varies in time with seasonal, annual and decadal variations, which could be linked to natural oscillations, further excludes the possibility that this pattern could be determined and/or dominated by cultural noise. In fact, if cultural noise had been the main driver of the daily oscillation, then one would have expected a more regular day-night pattern. There are no evidences known to us that cultural noise on and around the Mt. Vesuvius has changed similarly to natural oscillation during those years. Thus, the result of the analysis must manifest the actual natural evolution in the geophysical behavior of the volcano.

## 5. Conclusion

Using the BKE record, which is complete for events with magnitude $M \geq 0.2$, we have clearly demonstrated the existence of a 24-h cyclical modulation of the seismicity of the Mt. Vesuvius with maxima during the nighttime. For numerous years this seismic oscillation has been more enhanced during summer times than during winter times.

We have demonstrated that the OVO record, which is complete for events with magnitude $M \geq 1.9$, is inadequate to study this daily oscillation because using this record this pattern vanishes for numerous years. This result questions Duma and Vilardo (1998) and makes the present study necessary for properly investigating this phenomenon.

We have demonstrated that the adoption of the BKE record allows: 1) to solve the statistical issue supporting the existence of a daily oscillation in the seismic activity of the Mt. Vesuvius; 2) to find additional seasonal patterns (e.g. the daily oscillation is stressed during summer times). Thus, because the above properties, BKE can be compared with other variables which helps to better determine the physical mechanisms explaining this phenomenon. We suggested a number of mechanisms linked with the cooling/warming diurnal cycle of the volcanic edifice, as formerly reported for the Gran Sasso mountain by Paparo et al. (2003), and/or with the daily oscillation of the geomagnetic field.

In addition, the finding also suggests and provides a measure for monitoring the analyzed daily variation in the Mt. Vesuvius seismic activity: see Figure 10. The suggested index compares the number of seismic events occurring during nighttime (6:00 pm - 6:00 am) and daytime (6:00 am - 6:00 pm) for moving intervals that could be of 12 months to filter out the annual oscillation, and of 4 months to highlight it.

The proposed index manifests a rich dynamics which could be monitored to complement other methodologies currently adopted to determine the seismic status of the volcano and for hazard prevention. We also suggest that the Mt. Vesuvius, as well as any other active volcano, be monitored with thermal, magnetic and electric sensors.

Figure 1: [A] The seismic events recorded at the Osservatorio Vesuviano Ovest (OVO, 1972-2014) and [B] at the Bunker Est (BKE, 1999-2014) stations on the Mt. Vesuvius. The red line at M=1.9 and M=0.2 refers to the completeness level of the two records respectively.

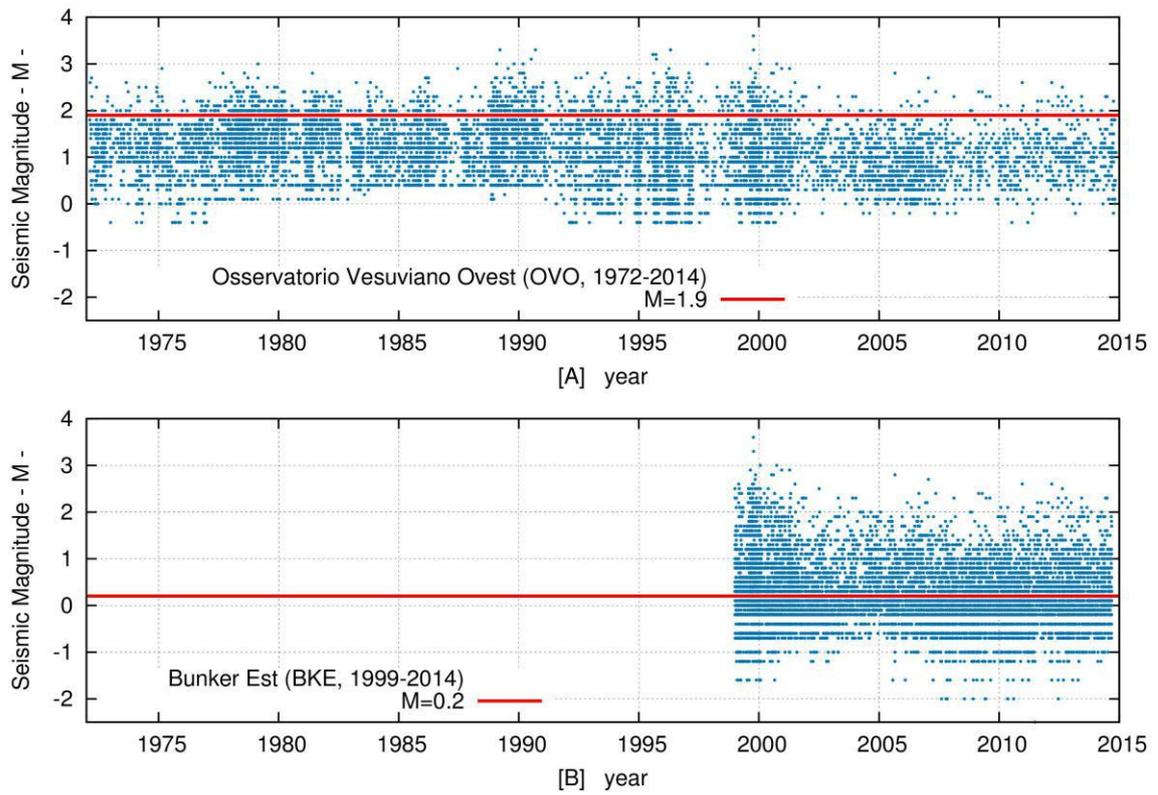

Figure 2: [A] Hourly histogram of the seismic data measured at the OVO station from 1972 to 1996. [B] Hourly histogram of the seismic data measured at the OVO station from 1997 to 2014. The sets contain only events with magnitude M $\geq$ 1.9 for which the catalog is complete. The gray shadow refers to the nighttime.

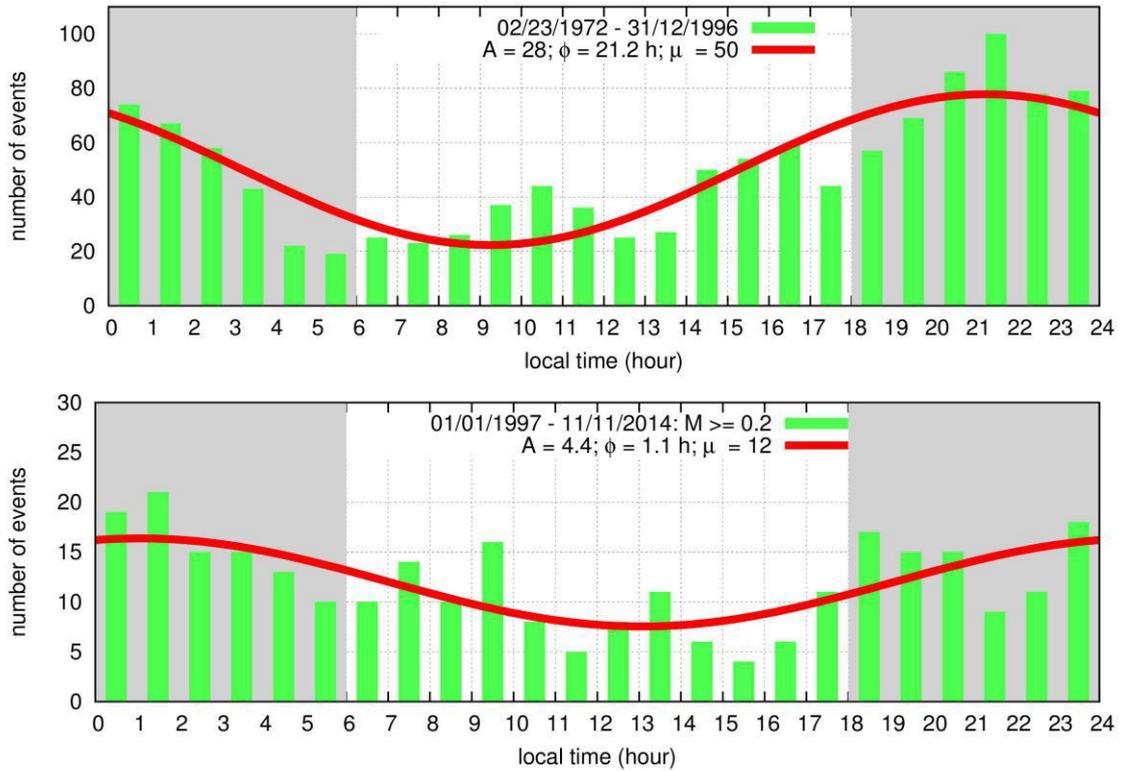

Figure 3: Sequence of 24-hour histogram of the seismic data measured at the OVO Vesuvius station for each year from 1972 to 2014. The set contains only events with magnitude M ≥ 1.9 for which the catalog is complete. The gray shadow refers to the nighttime.

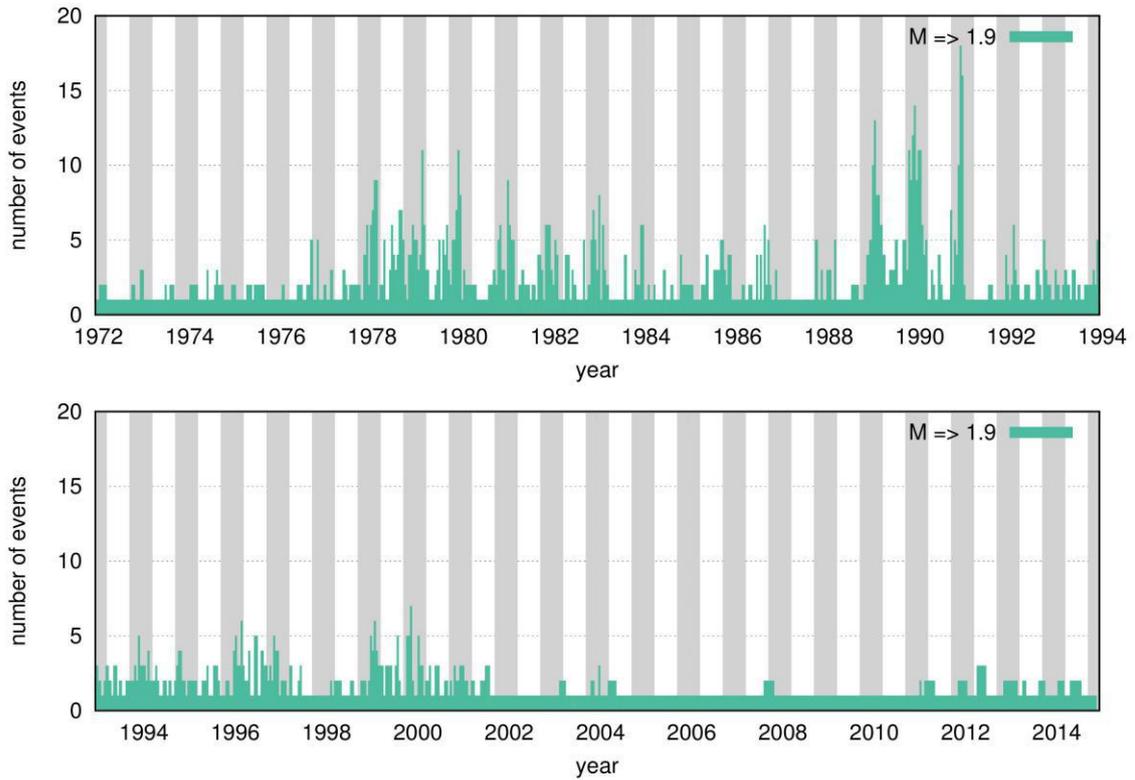

Figure 4: [A] Hourly histogram of the seismic data measured at the BKE Vesuvius station from 1999 to 2014. [B] Subset containing only events with magnitude M ≥ 0.2 for which the catalog is complete. The gray shadow refers to the nighttime.

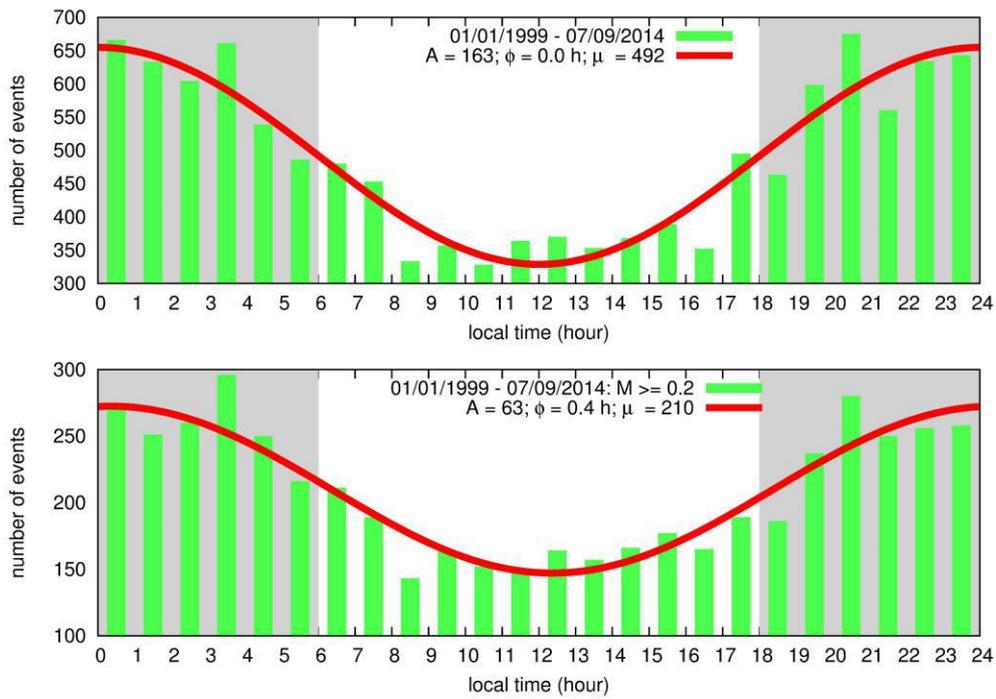

Figure 5:[A] Hourly histogram of the seismic data measured at the BKE Vesuvius station divided in three equal length subcatalogs. [B] Subsets containing only events with magnitude M ≥ 0.2 for which the catalog is complete. The gray shadow refers to the nighttime.

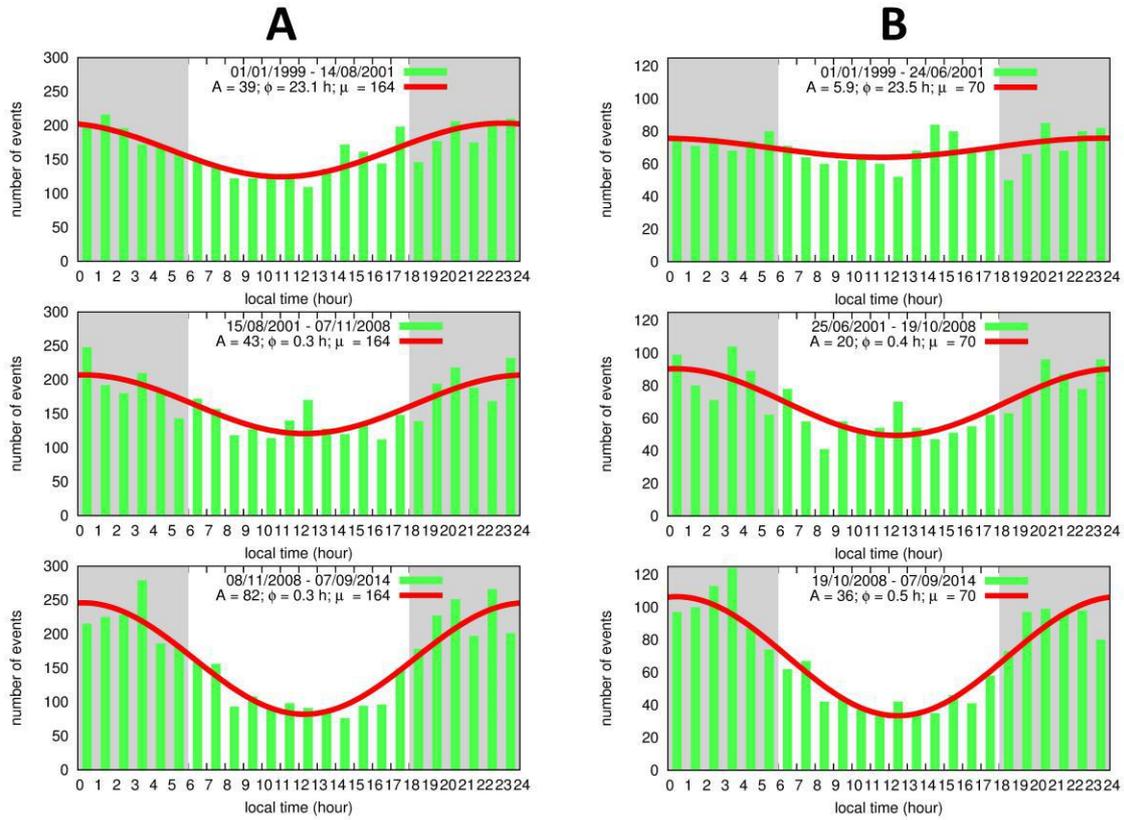

Figure 6:[A] Hourly histogram of the seismic data measured at the BKE Vesuvius station divided in D, E and J Lloyds seasons. [B] Subsets containing only events with magnitude M $\geq 0.2$ for which the catalog is complete. The gray shadow refers to the nighttime.

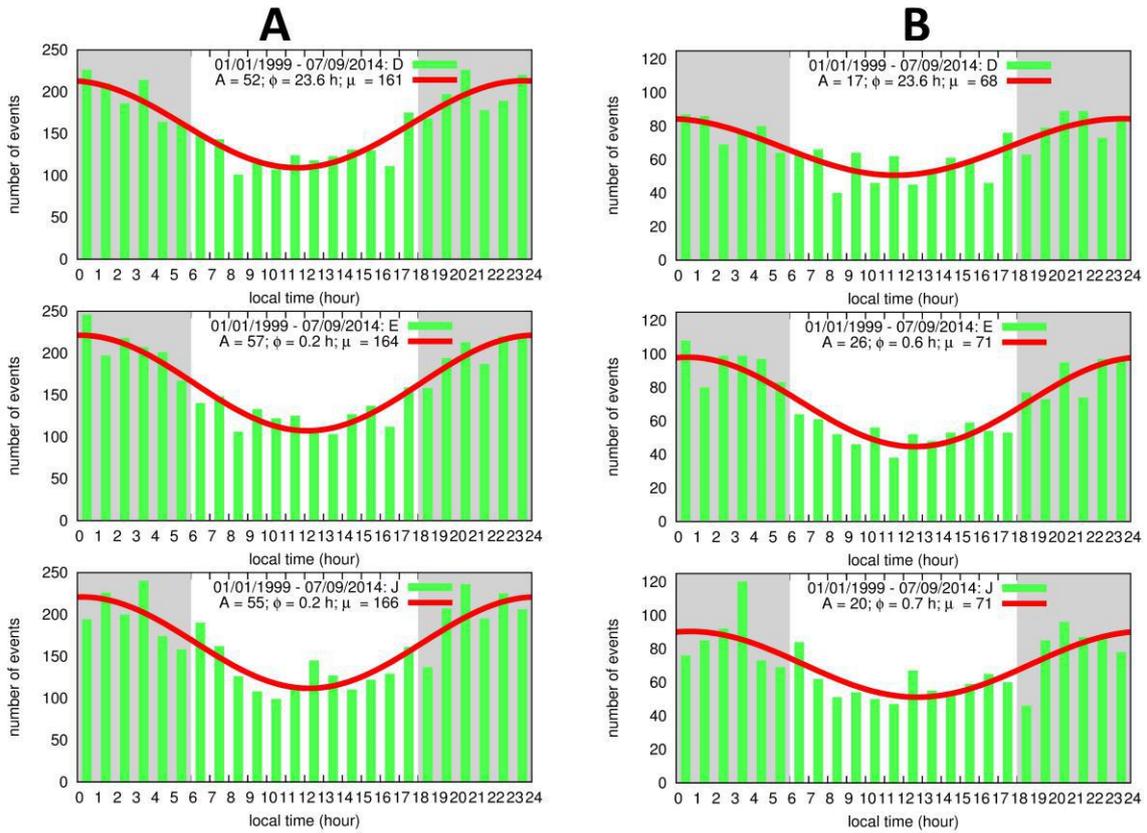

Figure 7:[A] Hourly histogram of the seismic data measured at the BKE Vesuvius station divided in D, E and J Lloyds seasons referring to the last third from 8/11/2008 to 7/9/2014. [B] Subsets containing only events with magnitude M ≥ 0.2 for which the catalog is complete. The gray shadow refers to the nighttime.

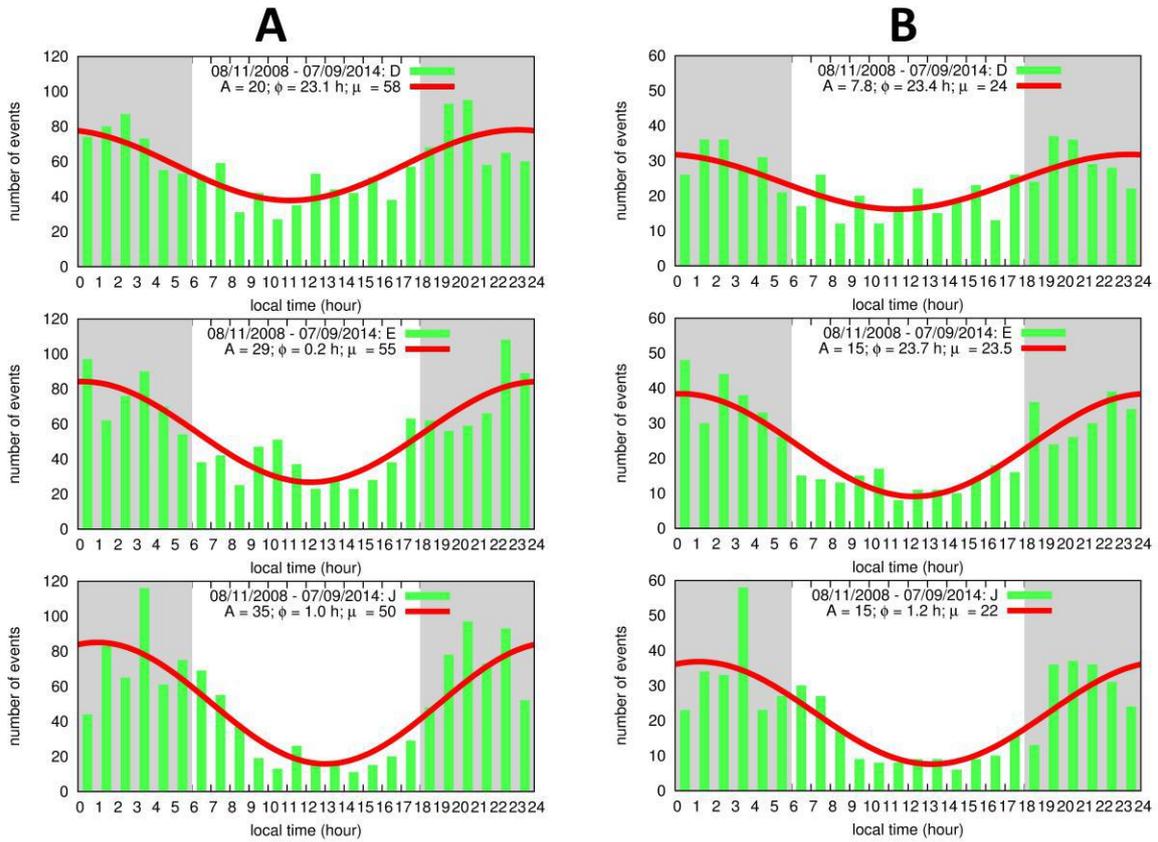

Figure 8: [A] Sequence of 24-hour histogram of the seismic data measured at the BKE Vesuvius station for each year from 1999 to 2014. [B] Subset containing only events with magnitude M ≥ 0.2 for which the catalog is complete. The gray shadow refers to the nighttime.

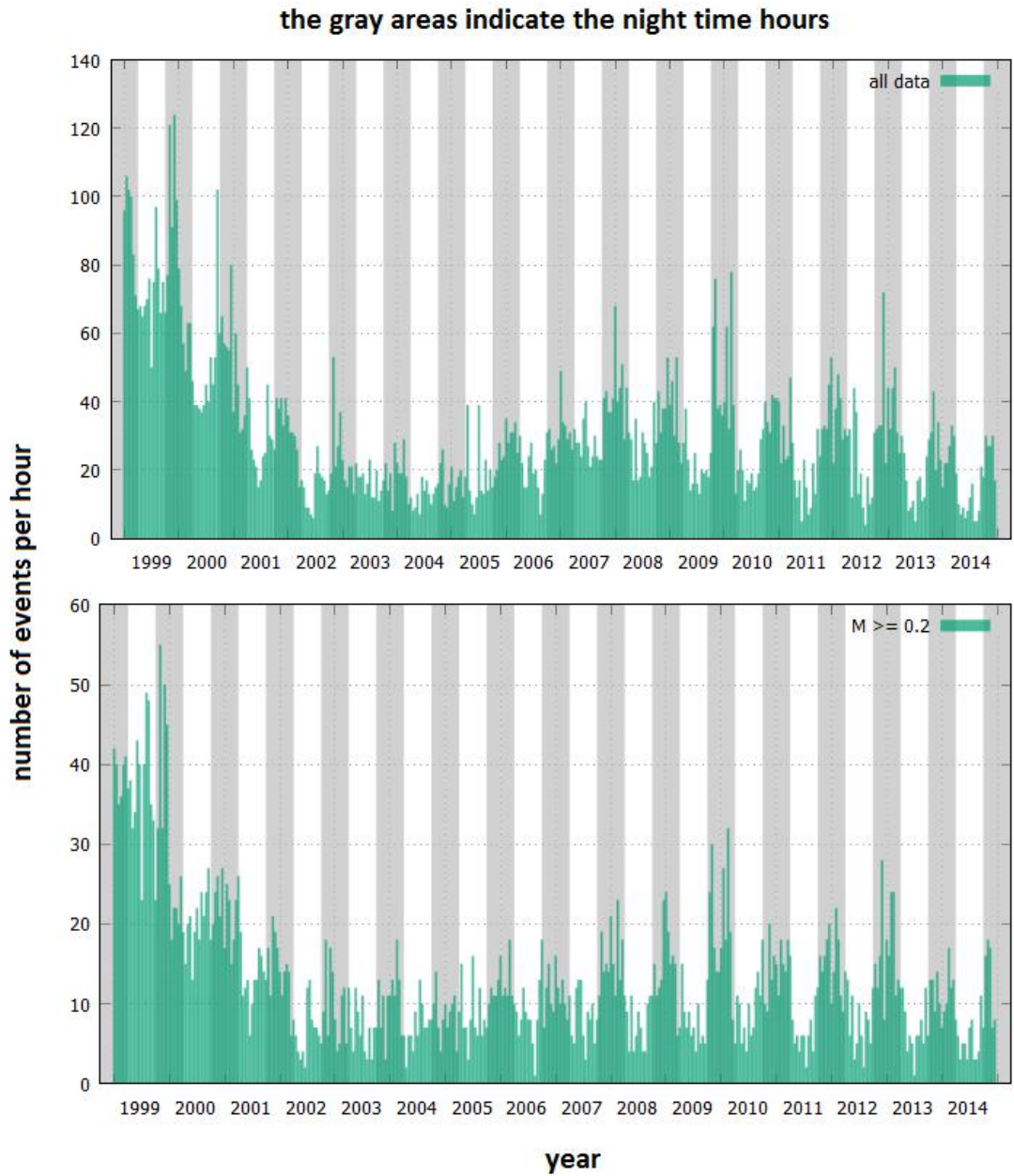

Figure 9: [A] Sequence of 24-hour histogram of the seismic data measured at the BKE Vesuvius station for each year from 1999 to 2014 for the winter D months. [B] and [C] like [A] for the equinoxes (E) and summer (J) months. The gray shadow refers to the nighttime.

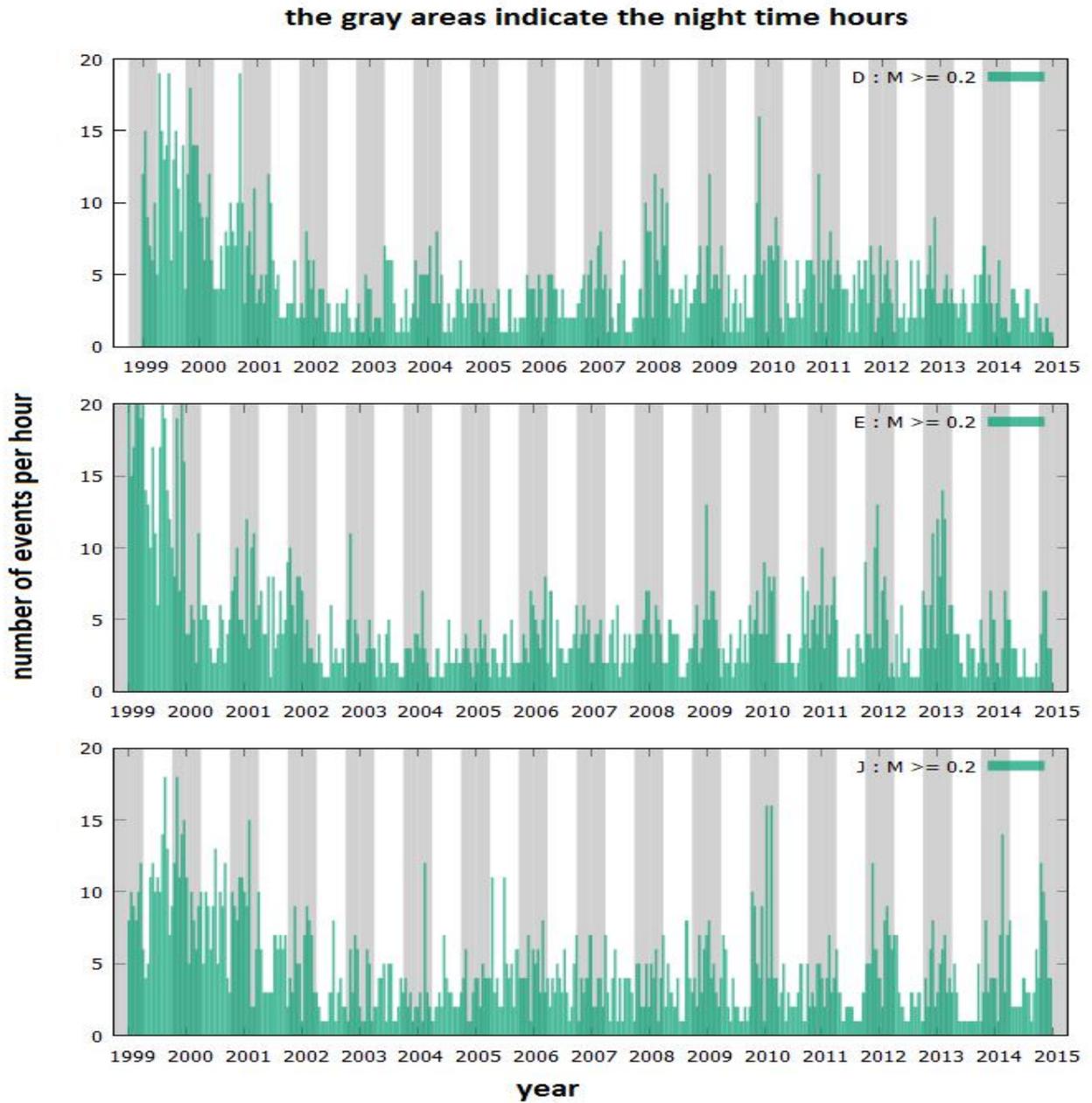

Figure 10: [A] Number of BKE seismic events for M ≥ 0.2 occurred during nighttime (6:00 pm - 6:00 am LT) and during daytime (6:00 pm - 6:00 am LT) for 12-month intervals. [B] As in [A] for 4-month intervals. The 12-month and 4-month intervals move by 1-month step. [C] $\chi^2$ curves relative to each interval in the two cases against the 95% and 99% confidence levels.

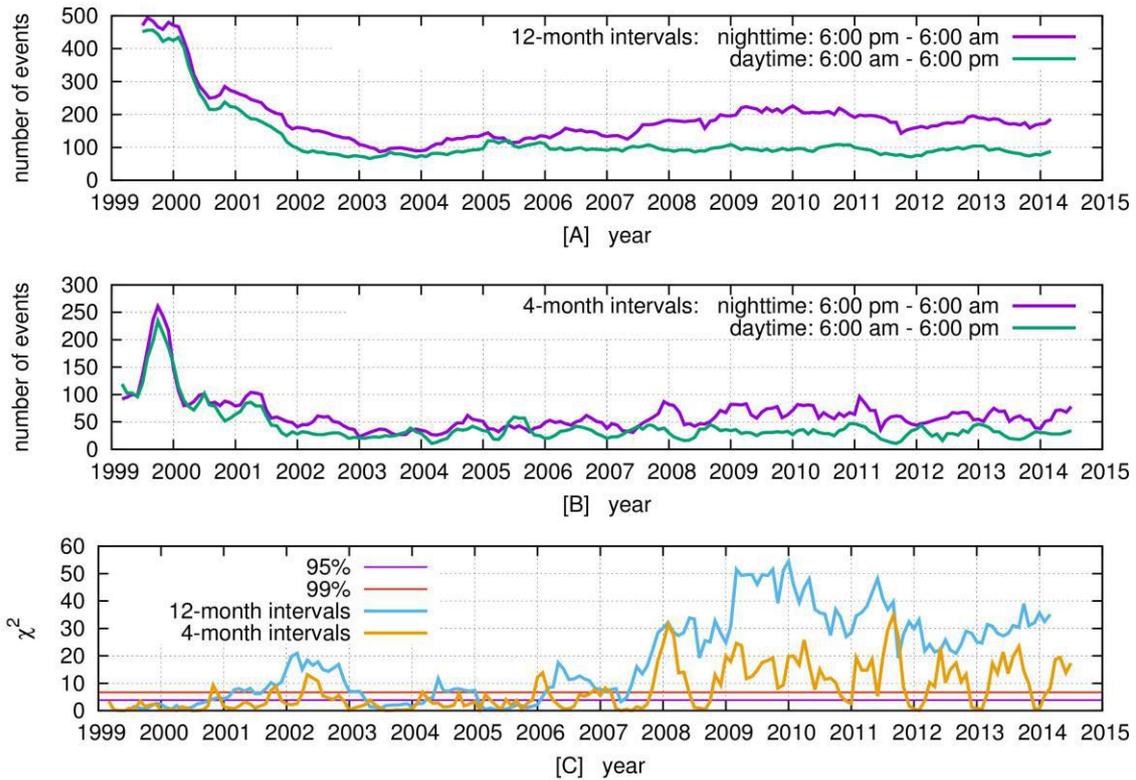

Table 1: The $\chi^2$ test with 23 degrees of freedom (df) when the hourly frequency is used and df = 1 degree of freedom.

|  | $\chi^2$ (hour) | $\chi^2$ (day/night) |
|---|---|---|
| df | 23 | 1 |
| 95% confidence | 35.2 | 3.84 |
| 99% confidence | 41.6 | 6.63 |
| all | 722 | 538 |
| all M $\geqslant$ 0.2 | 257 | 193 |
| 1 third | 146 | 71 |
| 2 third | 210 | 109 |
| 3 third | 582 | 454 |
| 1 third, M $\geqslant$ 0.2 | 29 | 3.1 |
| 2 third, M $\geqslant$ 0.2 | 108 | 61 |
| 3 third, M $\geqslant$ 0.2 | 265 | 211 |
| D | 238 | 170 |
| E | 268 | 208 |
| J | 274 | 162 |
| D, M $\geqslant$ 0.2 | 74 | 41 |
| E, M $\geqslant$ 0.2 | 143 | 114 |
| J, M $\geqslant$ 0.2 | 112 | 49 |
| 3 third, D | 139 | 79 |
| 3 third, E | 243 | 149 |
| 3 third, J | 430 | 258 |
| 3 third, D, M $\geqslant$ 0.2 | 57 | 30 |
| 3 third, E, M $\geqslant$ 0.2 | 136 | 106 |
| 3 third, J, M $\geqslant$ 0.2 | 180 | 83 |

Table 2: Diurnal cycle in the entire catalog of earthquakes.

|  | amplitude | vpe | Phase (°) | max of occurrence (h) | hourly mean | A/vpe (confidence level) |
|---|---|---|---|---|---|---|
| 11804 earthquakes (1/1/1999-7/9/2014) | 33.3 | 2.9 | 89.7 | 0.0 | 41.0 | 11.5 (> 99%) |
| 3935 earthquakes (1/1/1999-14/8/2001) | 24.3 | 4.2 | 103.3 | 23.1 | 13.7 | 5.8 (> 99%) |
| 3935 earthquakes (14/8/2001-7/11/2008) | 26.6 | 3.3 | 86.5 | 0.2 | 13.7 | 8.1 (> 99%) |
| 3934 earthquakes (8/11/2008- 7/9/2014) | 49.9 | 4.9 | 84.6 | 0.4 | 13.7 | 10.2 (> 99%) |

Table 3: Seasonal cycle computed starting from the entire catalog relative to 1/1/1999-7/9/2014 interval and containing 11804 earthquakes.

| number earthquakes | Lloyds season | amplitude | vpe | phase (°) | max of occurrence (h) | hourly mean | A/vpe (confidence level) |
|---|---|---|---|---|---|---|---|
| 3868 | D (J,F,N,D) | 32.4 | 2.3 | 95.3 | 23.6 | 13.4 | 14.1 (> 99%) |
| 3944 | E (M,A,S,O) | 35.0 | 4.2 | 88.3 | 0.1 | 13.7 | 8.3 (> 99%) |
| 3992 | J (M,J,J,A) | 32.8 | 4.2 | 87.9 | 0.1 | 13.9 | 7.8 (> 99%) |

Table 4: Seasonal cycle computed starting from the sub-catalog relative to the 8/1/2008-7/9/2014 interval and containing 3934 earthquakes.

| number earthquakes | Lloyds season | amplitude | vpe | phase (°) | max of occurrence (h) | hourly mean | A/vpe (confidence level) |
|---|---|---|---|---|---|---|---|
| 1391 | D (J,F,N,D) | 33.3 | 7.4 | 104.0 | 23.1 | 4.8 | 4.5 (> 99%) |
| 1332 | E (M,A,S,O) | 54.0 | 10.2 | 84.1 | 0.4 | 4.6 | 5.3 (> 99%) |
| 1211 | J (M,J,J,A) | 67.1 | 7.5 | 77.7 | 0.8 | 4.2 | 8.9 (> 99%) |

Table 5: Diurnal cycle in the catalogue of earthquakes with magnitude M ≥ 0.2.

| | amplitude | vpe | Phase (°) | max of occurrence (h) | hourly mean | A/vpe (confidence level) |
|---|---|---|---|---|---|---|
| 5035 earthquakes M ≥ 0.2 (1/1/1999-7/9/2014) | 30.5 | 2.4 | 84.2 | 0.4 | 17.5 | 11.5 (> 99%) |
| 1729 earthquakes (1/1/1999-14/8/2001) | 10.6 | 4.9 | 93.2 | 23.8 | 6.0 | 2.1 (95%) |
| 1651 earthquakes (14/8/2001-7/11/2008) | 28.7 | 2.9 | 84.5 | 0.4 | 5.7 | 9.9 (> 99%) |
| 1655 earthquakes (8/11/2008- 7/9/2014) | 54.4 | 4.8 | 81.7 | 0.5 | 5.7 | 11.3 (> 99%) |

Table 6: Seasonal cycle computed starting from the catalogue containing 5035 earthquakes with magnitude M ≥ 0.2 and relative to 1/1/1999-7/9/2014 interval.

| number earthquakes | Lloyds season | amplitude | vpe | phase (°) | max of occurrence (h) | hourly mean | A/vpe (confidence level) |
|---|---|---|---|---|---|---|---|
| 1622 | D (J,F,N,D) | 25.3 | 3.7 | 95.4 | 23.6 | 5.6 | 6.8 (> 99%) |
| 1714 | E (M,A,S,O) | 38.0 | 4.5 | 81.2 | 0.6 | 5.9 | 8.4 (> 99%) |
| 1699 | J (M,J,J,A) | 27.3 | 4.7 | 81.3 | 0.6 | 5.9 | 8.9 (> 99%) |

Table 7: Seasonal cycle computed starting from the catalogue containing 1655 earthquakes with magnitude M ≥ 0.2 and relative to 8/11/2008-7/9/2014 interval.

| number earthquakes | Lloyds season | amplitude | vpe | phase (°) | max of occurrence (h) | hourly mean | A/vpe (confidence level) |
|---|---|---|---|---|---|---|---|
| 565 | D (J,F,N,D) | 32.9 | 6.5 | 94.3 | 23.7 | 2.0 | 5.1 (> 99%) |
| 557 | E (M,A,S,O) | 65.7 | 10.5 | 84.4 | 0.4 | 1.9 | 6.3 (> 99%) |
| 533 | J (M,J,J,A) | 62.7 | 9.9 | 74.8 | 1.0 | 1.9 | 6.3 (> 99%) |

Table 8: The amplitude in nanotesla (nT), phase in degree (º) and time maximum in hour (h) of the vertical (Z) component of the measured geomagnetic field in Central Italy.

| Station | component | number of days | amplitude (nT) | phase (°) | time maximum of occurrence (h) |
|---|---|---|---|---|---|
| L'Aquila | | | | | |
| | Z (total) | 3354 | 6.04 | 108.5 | 22.8 |
| | Z (D months) | 1085 | 3.24 | 126.4 | 21.6 |
| | Z (E months) | 1118 | 6.90 | 104.4 | 23.0 |
| | Z (J months) | 1151 | 8.01 | 105.2 | 23.0 |
| Capri | | | | | |
| | Z (total) | 2145 | 6.95 | 110.8 | 22.6 |
| | Z (D months) | 706 | 4.14 | 119.8 | 22.0 |
| | Z (E months) | 731 | 7.89 | 105.1 | 23.0 |
| | Z (J months) | 708 | 8.88 | 112.1 | 22.5 |

Table 9: The amplitude in nanotesla (nT), phase in degree (º) and time maximum in hour (h) of the vertical (Z) component of the measured geomagnetic field in Spain and Turkey.

| Station | component | number of days | amplitude (nT) | vpe | phase (°) | maximum of occurrence (h) |
|---|---|---|---|---|---|---|
| San Fernando (Spain) | Z (total) | 1826 | 7.1 | 0.1 | 99 | 23.4 |
| | Z (D months) | 601 | 5.0 | 0.1 | 109 | 22.7 |
| | Z (E months) | 610 | 7.5 | 0.2 | 102 | 23.2 |
| | Z (J months) | 615 | 9.4 | 0.1 | 101 | 23.3 |
| Ebro (Spain) | Z (total) | 1461 | 6.0 | 0.1 | 97 | 23.5 |
| | Z (D months) | 481 | 3.5 | 0.1 | 104 | 23.1 |
| | Z (E months) | 488 | 6.4 | 0.2 | 95 | 23.7 |
| | Z (J months) | 492 | 8.0 | 0.1 | 97 | 23.5 |
| San Pablo (Spain) | Z (total) | 1065 | 5.9 | 0.1 | 102 | 23.2 |
| | Z (D months) | 361 | 3.8 | 0.1 | 121 | 21.9 |
| | Z (E months) | 366 | 6.4 | 0.1 | 100 | 23.3 |
| | Z (J months) | 338 | 8.0 | 0.1 | 94 | 23.7 |
| Istanbul-Kandilli (Turkey) | Z (total) | | 6.4 | 0.0 | 93 | 23.8 |
| | Z (D months) | | 3.4 | 0.0 | 110 | 23.3 |
| | Z (E months) | | 7.1 | 0.0 | 90 | 0.9 |
| | Z (J months) | | 8.8 | 0.0 | 88 | 0.9 |
| Iznik (Turkey) | Z (total) | | 4.2 | 0.0 | 81 | 0.6 |
| | Z (D months) | | 1.9 | 0.1 | 100 | 23.3 |
| | Z (E months) | | 4.7 | 0.1 | 77 | 0.9 |
| | Z (J months) | | 6.1 | 0.1 | 76 | 0.9 |